\documentclass[notitlepage,aps,pra]{revtex4-2}

\usepackage{physics}
\usepackage{amsmath}
\usepackage{xcolor}
\usepackage{graphicx}
\usepackage[hidelinks]{hyperref}
\usepackage{cleveref}
\usepackage{float}
\usepackage[normalem]{ulem} 
\crefname{equation}{Eq.}{} 

\newcommand{\pati}[2]{}

\newcommand{\kdk}[1]{}

\begin{document}

\title{Dispersion-Mediated Space-Time States}

\author{Klaas De Kinder}
\author{Christophe Caloz}
\email[]{christophe.caloz@kuleuven.be}
\affiliation{Department of Electrical Engineering, KU Leuven, Leuven, 3000,  Belgium}

\date{\today}

\begin{abstract}
    Space-time varying media enable unprecedented control over electromagnetic waves, yet most existing studies assume idealized, nondispersive materials and thus fail to capture the intrinsic frequency dispersion of realistic platforms. Here, we develop a general framework for dispersive space-time varying systems that rigorously identifies the physically allowed frequency transitions of waves scattered at moving interfaces. Unlike previous approaches, our method is valid for arbitrary dispersion profiles, including resonances, and does not rely on the commonly used frame hopping approach, allowing treatment of multiple-velocity and accelerated systems. Applying this framework to canonical Drude and Lorentz media, we uncover a family of dispersion-mediated space-time states that arise from the multiple frequency transitions permitted by material dispersion. These states extend beyond conventional nondispersive scattering, revealing qualitatively new regimes of space-time scattering behavior. Beyond the frequency transitions, we derive the scattering coefficients for dispersive moving interfaces and provide a Fourier-domain formulation that yields a complete electromagnetic scattering solution for both monochromatic waves and broadband pulses. Our results establish a rigorous foundation for the design of realistic space-time metamaterials, with immediate relevance to emerging experiments in epsilon-near-zero optical platforms and open pathways for dispersion-engineered wave manipulation.
\end{abstract}

\maketitle

\section{Introduction}

    \pati{Definition Space-Time Varying Systems}{
    }

    Space-time varying media are advanced structures whose material properties, such as the refractive index, are modulated in both space and time~\cite{Cassedy1963_PUB,Cassedy1967_PUB,Caloz2019a_ST_Metamaterials_PUB,Caloz2019b_ST_Metamaterials_PUB}, with the modulations typically produced by an external drive such as electronic, optical, acoustic or mechanical pumps~\cite{Saleh2019_BOOK,Rhodes1981_Signal_Proc_PUB,Shalaev2019_Active_Meta_PUB}. Compared to static media, space-time varying media support qualitatively different wave interactions and enable new physical effects~\cite{Caloz2022_GSTEMs_PUB}, which in turn unlock a range of novel applications, including motion-induced photon cooling~\cite{Pendry2024_Air_Cond_Phot_USTEM_PUB}, arbitrary pulse shaping~\cite{Bahrami2025_Pulse_Shap_ASTEM_PUB}, generalized frequency chirping~\cite{DeKinder2025_Scat_Chirp_ASTEM}, interface-based light amplification~\cite{Pendry2021_Gain_PUB,Luo2023_PUB}, time reversal~\cite{Fink2016_TR_TEM_PUB}, temporal beam splitting~\cite{Guerreiro2003_PUB} and Doppler pulse amplification~\cite{DeKinder2025_DoPA_USTEM_PUB}.

    \pati{Importance of Dispersion}{
     }

     While many of these applications have been explored in idealized, nondispersive media, physical materials are inherently causal and exhibit therefore frequency-dependent dispersive responses that can strongly influence wave propagation, causing effects such as pulse distortion and frequency chirping~\cite{Saleh2019_BOOK}. Moreover, dispersion is particularly relevant in experiments, where it may dominate the system behavior over the idealized, nondispersive case. Recent works have begun addressing dispersive \emph{time}-varying media~\cite{Fante1971_Transmission_TEM_PUB,Fante1973_Prop_EM_TEM_PUB,Pendry2022_Review_TEM_PUB,Engheta2025_Disp_Hyper_Media_TEM_PUB,Horsley2023_Eigenpulses_PUB,Engheta2021_Dispersion_PUB,Mirmoosa2022_Dipole_PUB,Joannopoulos2024_Opt_Prop_PUB,Maslov2021_Light_Scat_PUB,Maslov2021_Const_Rel_PUB,Maslov2024_Dispersion_PUB,Monticone2022_Dispersion_PUB}, deriving new fundamental bounds, such as Kramers-Kronig relations~\cite{Kramers1927_KK_Relations_PUB,Kronig1926_KK_Relations_PUB} for time-varying media and exploring their impact on wave dynamics~\cite{Monticone2021_Spectral_Caus_PUB,Engheta2021_Kramers_Kronig_PUB}.

    \pati{Gap}{
    }

    Despite these advances, dispersive \emph{space-time} varying systems have received comparatively little attention. Apart from a series of studies by the group of Agrawal~\cite{Agrawal2015_Temporal_Analog_USTEM_PUB,Agrawal2016_Spectral_Splitting_PUB,Agrawal2021_Refl_and_Trans_USTEM_PUB}, most existing analyses are limited to purely temporal modulations. Moreover, the approaches in~\cite{Agrawal2015_Temporal_Analog_USTEM_PUB,Agrawal2016_Spectral_Splitting_PUB,Agrawal2021_Refl_and_Trans_USTEM_PUB} are limited by several assumptions: (1)~polynomial Taylor approximation of the dispersion relations, (2)~slowly varying envelope approximation and (3)~frame hopping to a comoving reference frame. These assumptions significantly limit their range of applicability, especially near resonances or in systems with multiple or nonuniform velocities. Meanwhile, experimental interest in space-time varying systems is rapidly growing, with $\epsilon$-near-zero (ENZ) materials emerging as a particularly promising platform in the optical regime~\cite{Shalaev2023_PTC_TEM_PUB,Boyd2020_Time_Refr_ENZ_TEM_PUB,Segev2023_Single_Cycle_TEM_PUB,Kinsey2025_ST_Knife_TEM_PUB}. Because ENZ materials are highly dispersive, accurately predicting the response of a moving modulation requires a systematic framework capable of handling fully dispersive space-time varying systems, including frequencies near resonance.
      
    \pati{Contribution}{
    }

    In this work, we provide such a framework by identifying the physically allowed frequency transitions of scattered waves at a uniformly moving interface separating two dispersive media, opening the pathway for dispersion-engineered wave manipulation. We show that dispersion can create additional frequency-transition branches that have no counterpart in the nondispersive limit, giving rise to new ``dispersion-mediated space-time states'', depending on the dispersive properties of the media, the interface velocity and the incident frequency. Our approach is valid for arbitrary dispersion profiles, including resonant regimes where the Taylor expansion of the dispersion relation break down and does not rely on frame hopping, allowing for multiple-velocity systems~\cite{Deck-Leger2019_Uni_Vel_PUB,Bahrami2025_Wedges_USTEM_PUB,Luo2025_ST_Super_Focu_USTEM_PUB} and arbitrarily accelerated interfaces~\cite{DeKinder2025_Scat_Chirp_ASTEM}. Beyond identifying the spectral content of the scattered fields, we further derive the scattering coefficients for dispersive moving interfaces in the subluminal regime and generalize these results to broadband pulses using a Fourier-domain formulation. This allows us to obtain a complete electromagnetic scattering solution that consistently incorporates both interface motion and material dispersion, enabling quantitative predictions for realistic experimental platforms such as epsilon-near-zero materials in the optical regime.
      
    \pati{Organization}{
    }

    The paper is structured as follows. Section~\ref{sec:Frequency_Transitions} derives general expressions for the frequency transitions of the scattered waves and introduces two criteria to select the physically meaningful solutions. Section~\ref{sec:Application_to_Drude_Media} applies this methodology to a lossless Drude model and identifies ten distinct scattering regimes defined by the type and number of scattered waves. Section~\ref{sec:Application_to_Lorentz_Media} extends the analysis to the Lorentz model, revealing $21$ different scattering regimes. Section~\ref{sec:Scattering_Coefficients} derives the scattering coefficients in the subluminal regime for a monochromatic excitation while Sec.~\ref{sec:Pulses} generalizes these results to broadband pulses, providing a full solution to the dispersive electromagnetic scattering problem. Finally, Sec.~\ref{sec:Conclusions} summarizes our findings and presents concluding remarks.

\section{Frequency Transitions}\label{sec:Frequency_Transitions}
    \pati{Assumptions}{
    }

    We study the problem of electromagnetic scattering at a modulation interface moving with uniform velocity $v_{\text{m}}$ along the $z$-axis and separating two isotropic, linear, passive but dispersive media~\footnote{For simplicity, the media are also considered to be lossless. However, loss may be straightforwardly added to the dispersive responses.}. Both media are characterized by frequency-dependent refractive indices, $n_{1,2}{\left[\omega\right]}$, which satisfy the dispersion relation in their related medium: $\beta_{1,2}^{2}{\left[\omega\right]} = n_{1,2}^{2}{\left[\omega\right]}\omega^{2}/c^{2}$. An incident pulse in the first medium impinges on the moving interface and generates reflected and transmitted waves whose frequencies differ from the incident frequency due to the combined effects of Doppler shift and material dispersion.

    \pati{Phase Matching Method}{
    }

    The spectral components of the scattered waves are obtained by imposing phase matching~\cite{Deck-Leger2024_Dispersion_CONF} between the scattered and the incident waves along the trajectory of the interface. Writing the fields with the time-harmonic convention $\exp\left(i\left(\beta{\left[\omega\right]}z -\omega t\right)\right)$, phase continuity along $z = v_{\text{m}}t$ yields
    \begin{subequations}\label{eq:Phase_Matching_Conditions}
    \begin{align}
        \beta_{1}{\left[\omega_{1}\right]}z - \omega_{1}t &= \left.\beta_{1}{\left[\omega_{\text{i}}\right]}z - \omega_{\text{i}}t\right|_{z=v_{\text{m}}t}\,, \\
        \beta_{2}{\left[\omega_{2}\right]}z - \omega_{2}t &= \left.\beta_{1}{\left[\omega_{\text{i}}\right]}z - \omega_{\text{i}}t\right|_{z=v_{\text{m}}t}\,,
    \end{align}
    \end{subequations}
    where $\omega_{\text{i}}$, $\omega_{1}$ and $\omega_{2}$ are the incident, reflected and transmitted frequencies, respectively. Inserting the dispersion relation, the trajectory parametrization and eliminating the explicit time-dependence in Eqs.~\eqref{eq:Phase_Matching_Conditions}, we obtain the following characteristic equations for the frequency transitions as a function of the incident frequency:
    \begin{subequations}\label{eq:General_Characterstic_Equations}
        \begin{align}
            \left(1+n_{1}{\left[\omega_{1}\right]}\frac{v_{\text{m}}}{c}\right)\omega_{1} &= \left(1-n_{1}{\left[\omega_{\text{i}}\right]}\frac{v_{\text{m}}}{c}\right)\omega_{\text{i}}\,, \\
            \left(1-n_{2}{\left[\omega_{2}\right]}\frac{v_{\text{m}}}{c}\right)\omega_{2} &= \left(1-n_{1}{\left[\omega_{\text{i}}\right]}\frac{v_{\text{m}}}{c}\right)\omega_{\text{i}}\,.
        \end{align}
    \end{subequations}
    In the nondispersive limit, i.e., $n_{1,2}{\left[\omega\right]} = n_{1,2}$, these equations reduce to the standard reflected and transmitted Doppler shifts at a constant moving interface~\cite{Caloz2019b_ST_Metamaterials_PUB}. However, in the dispersive case, Eqs.~\eqref{eq:General_Characterstic_Equations} can admit multiple solutions for each medium, each corresponding to a distinct scattered mode. Therefore, we denote the frequency of the $j$-th scattered wave in medium $i$ by $\omega_{i}^{\pm(j)}$, where $\pm$ indicates forward ($+$) or backward ($-$) propagation and $j=1,2,\dots$ labels distinct scattered waves.

    \pati{Physical Wave Conditions}{
    }

    Not all the mathematical solutions of Eqs~\eqref{eq:General_Characterstic_Equations} correspond to physically meaningful waves. To select the physical frequency transitions, we enforce two conditions: group-velocity restriction and media passivity. The group-velocity restriction stems from kinematic considerations and states that a scattered wave must propagate in such a way that it is not overtaken by the moving interface. If the interface moves as $z = v_{\text{m}}t$ and a wavepacket as $z=v_{\text{g}}t$, where $v_{\text{g}}$ is the group velocity, the wave remains physically meaningful only if its trajectory is appropriately located ahead of or behind the interface, depending on which medium it propagates in. Additionally, each scattered wave experiences a frequency shift governed by Eqs.~\eqref{eq:General_Characterstic_Equations}. Because the group velocity is frequency dependent, $v_{\text{g},i}{\left[\omega_{i}^{\pm(j)}\right]}$ generally differs from that of the incident wave, $v_{\text{g},1}{\left[\omega_{\text{i}}\right]}$. Therefore, the group-velocity restriction must be applied individually to each scattered frequency. For waves scattered in the second medium, physical modes must satisfy $v_{\text{g},2}{\left[\omega_{2}^{\pm(j)}\right]} > v_{\text{m}}$, ensuring that the transmitted wave outruns the interface. For example, assuming wave propagation in the $+z$-direction, a codirectional interface ($v_{\text{m}} >0$) generating a forward-propagating transmitted wave ($\omega_{2}^{+(1)}$), must allow the wave to propagate faster than the interface; otherwise, the interface would immediately catch up with it and annihilate it. Similarly, for waves in the first medium, the restriction requires $v_{\text{g},1}{\left[\omega_{1}^{\pm(j)}\right]} < v_{\text{m}}$. The second constraint, media passivity, excludes solutions producing unphysical temporal or spatial amplification. Its precise form depends on the time-harmonic convention. With the convention $\exp\left(i\left(\beta{\left[\omega\right]}z - \omega t\right)\right)$ used here, physical \emph{forward}-propagating modes satisfy $\Im\omega < 0$ and $\Im\beta > 0$, whereas physical \emph{backward}-propagating modes satisfy $\Im \omega < 0$ and $\Im\beta < 0$. Table~\ref{tab:Summary_Conditions_Frequency_Transitions} summarizes the two conditions used to select the physical frequency transitions among the solutions of Eqs.~\eqref{eq:General_Characterstic_Equations}.

    \begin{table}[h!]
        \centering
        \renewcommand{\arraystretch}{2.0} 
        \setlength{\tabcolsep}{10pt} 
        \caption{Group-velocity restriction and media passivity conditions for physical frequency transitions.}
        \begin{tabular}{|c|c|c|}
        \hline
        Medium & Forward ($+$) & Backward ($-$) \\
        \hline
        1 & $v_{\text{g},1}{\left[\omega_{1}^{+(j)}\right]} < v_{\text{m}},  \Im\beta{\left[\omega_{1}^{+(j)}\right]} > 0$ & $v_{\text{g},1}{\left[\omega_{1}^{-(j)}\right]} < v_{\text{m}}, \Im\beta{\left[\omega_{1}^{-(j)}\right]} < 0$ \\
        \hline
        2 & $v_{\text{g},2}{\left[\omega_{2}^{+(j)}\right]} > v_{\text{m}}, \Im\beta{\left[\omega_{2}^{+(j)}\right]} > 0$ & $v_{\text{g},2}{\left[\omega_{2}^{-(j)}\right]} > v_{\text{m}}, \Im\beta{\left[\omega_{2}^{-(j)}\right]} < 0$ \\
        \hline
        \multicolumn{3}{|c|}{\begin{tabular}{c}
        $\Im\omega_{i}^{\pm(j)} < 0$ for all waves \\
        \end{tabular}} \\
        \hline
        \end{tabular}
        \label{tab:Summary_Conditions_Frequency_Transitions}
    \end{table}

    \section{Application to Drude Media}\label{sec:Application_to_Drude_Media}
        \subsection{Specific Drude Solutions}
            \pati{Frequency Transitions Solutions}{
            }

\begin{figure}[h!]
                \centering
                \includegraphics[width=1.0\linewidth]{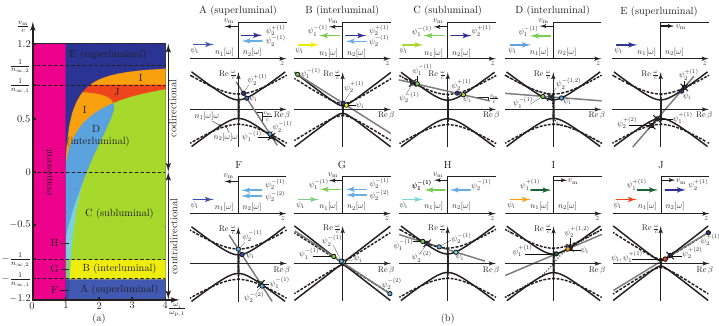}
                \caption{Dispersion-mediated space-time states for the lossless Drude model [Eq.~\eqref{eq:Refractive_Index_Profile_Free_Electron_Plasma_Model}] with $n_{\infty,1} = 1$, $\omega_{\text{p},1} = 5$, $n_{\infty,2} = 1.5$ and $\omega_{\text{p},2} = 10$, computed according to the conditions in Tab.~\ref{tab:Summary_Conditions_Frequency_Transitions}. (a)~Scattering regimes of the space-time states as a function of the modulation velocity, $v_{\text{m}}$, and the incident frequency, $\omega_{\text{i}}$. The dashed lines indicate the nondispersive velocity limits. (b)~Space-index perspective for each space-time state with corresponding spectral transition diagrams, where some of the solutions of Eq.~\eqref{eq:Frequency_Transitions_Example} are crossed out since they do not satisfy the conditions in Tab~\ref{tab:Summary_Conditions_Frequency_Transitions}.}
                \label{fig:Number_of_Waves}
            \end{figure}

            Let us apply the general methodology of Sec.~\ref{sec:Frequency_Transitions} to the lossless Drude model, in which the refractive index profiles of the two media are given by
            \begin{subequations}\label{eq:Refractive_Index_Profile_Free_Electron_Plasma_Model}
                \begin{align}
                    n_{1}{\left[\omega\right]} &= \sqrt{n_{\infty,1}^{2} - \frac{\omega_{\text{p},1}^{2}}{\omega^{2}}} \,, \\
                    n_{2}{\left[\omega\right]} &= \sqrt{n_{\infty,2}^{2} - \frac{\omega_{\text{p},2}^{2}}{\omega^{2}}} \,, 
                \end{align}
            \end{subequations}
            where $n_{\infty,1,2}$ are the high-frequency refractive indices and $\omega_{\text{p},1,2}$ are the plasma frequencies. We consider the general scenario in which both the high-frequency refractive index and the plasma frequency may be simultaneously modulated. Inserting Eqs.~\eqref{eq:Refractive_Index_Profile_Free_Electron_Plasma_Model} into Eqs.~\eqref{eq:General_Characterstic_Equations} yields two second-order polynomials in $\omega_{1,2}$, which can be easily solved as
            \begin{subequations}\label{eq:Frequency_Transitions_Example}
                \begin{align}
                    &\omega_{1} = \frac{1}{1-n_{\infty,1}^{2}v_{\text{m}}^{2}/c^{2}}\left(\left(1-n_{1}{\left[\omega_{\text{i}}\right]}v_{\text{m}}/c\right)\omega_{\text{i}} \pm v_{\text{m}}/c\sqrt{n_{\infty,1}^{2}\left(1-n_{1}{\left[\omega_{\text{i}}\right]}v_{\text{m}}/c\right)^{2}\omega_{\text{i}}^{2}-\left(1-n_{\infty,1}^{2}v_{\text{m}}^{2}/c^{2}\right)\omega_{\text{p},1}^{2}}\right) \,, \\
                    &\omega_{2} = \frac{1}{1-n_{\infty,2}^{2}v_{\text{m}}^{2}/c^{2}}\left(\left(1-n_{1}{\left[\omega_{\text{i}}\right]}v_{\text{m}}/c\right)\omega_{\text{i}} \pm v_{\text{m}}/c\sqrt{n_{\infty,2}^{2}\left(1-n_{1}{\left[\omega_{\text{i}}\right]}v_{\text{m}}/c\right)^{2}\omega_{\text{i}}^{2}-\left(1-n_{\infty,2}^{2}v_{\text{m}}^{2}/c^{2}\right)\omega_{\text{p},2}^{2}}\right) \,.
                \end{align}
            \end{subequations}
            Equations~\eqref{eq:Frequency_Transitions_Example} always yield two candidates for the reflected solutions and two candidates for the transmitted solutions, as expected from the second-order characteristic polynomial [Eq.~\eqref{eq:General_Characterstic_Equations}]. To sort out the physical solutions from these four candidate solutions, we apply the conditions summarized in Table~\ref{tab:Summary_Conditions_Frequency_Transitions}. Depending on the incident frequency ($\omega_{\text{i}}$), the modulation velocity ($v_{\text{m}}$) and dispersion parameters ($n_{\infty,1,2}$ and $\omega_{\text{p},1,2}$), the number and type of scattered waves vary. We refer to these distinct sets of allowed scattered waves as ``dispersion-mediated space-time states''. Figure~\ref{fig:Number_of_Waves} illustrates these states for the lossless Drude model [Eq.~\eqref{eq:Refractive_Index_Profile_Free_Electron_Plasma_Model}], with Fig.~\ref{fig:Number_of_Waves}a demonstrating the different scattering regimes as a function of the modulation velocity and the incident frequency and Fig.~\ref{fig:Number_of_Waves}b providing a space-index perspective and related spectral transition diagrams for each space-time state. Overall, ten different regimes, labeled A through J, are identified.

        \subsection{Conventional Modes}
            \pati{Regions~A--E}{
            }

            First, we identify the regions corresponding to known nondispersive scattering regimes~\cite{DeKinder2025_Scat_Chirp_ASTEM}, which are also recovered at high frequencies where dispersion becomes negligible. Region~C in Fig.~\ref{fig:Number_of_Waves} corresponds to the conventional subluminal regime, where an incident wave ($\psi_{\text{i}}$) generates a backward reflected ($\psi_{1}^{-(1)}$) and a forward transmitted wave ($\psi_{2}^{+(1)}$). No backward transmitted wave ($\psi_{2}^{-(1)}$) appears because such a solution would violate the group-velocity restriction: the corresponding wave would propagate slower (larger in absolute value) than the interface and would therefore be overtaken and annihilated, rendering the solution unphysical. In the spectral transition diagram, the group-velocity restriction is visualized by comparing the slope of the dispersion curve at the intersection point, which corresponds to the group velocity, with the slope of the transition line, which corresponds to the modulation velocity. Region~B corresponds to the contradirectional interluminal regime, where an incident wave produces reflected, later-backward and transmitted waves. Only in this regime are all the solutions in Eq.~\eqref{eq:Frequency_Transitions_Example} physically valid, as they all satisfy both the group-velocity and passivity conditions. Region~D represents the codirectional interluminal regime, in which the incident wave produces only a reflected wave. In contrast to the nondispersive case, this regime can also occur for contradirectional interfaces ($v_{\text{m}} < 0$). This regime can be seen as total internal reflection introduced in~\cite{Agrawal2015_Temporal_Analog_USTEM_PUB}. Finally, regions~A and~E correspond to superluminal regimes. In the codirectional superluminal regime (region~E), there is no scattering as the interface moves too fast for the incident wave to catch up. In the contradirectional superluminal regime (region~A), an incident wave generates a later-forward and a later-backward wave. There is no forward scattered wave in the first medium ($\psi_{1}^{-(1)}$) due to the group-velocity restriction. An important difference with the nondispersive limit is that the boundaries between these regimes become frequency dependent due to group-velocity dispersion. In nondispersive systems, the regime boundaries depend only on the modulation velocity. Dispersion shifts and reshapes these boundaries and even extends some of the regimes beyond their nondispersive counterparts, as for the subluminal region~C, codirectional interluminal region~D and codirectional superluminal region~E. 

        \subsection{Unconventional Modes}
            \pati{General Observation}{
            }

            In addition to the conventional, nondispersive-like regimes A--E, we uncover a set of purely dispersive space-time modes, which have no analog in nondispersive systems. These new modes correspond to regions~F through~J in Fig.~\ref{fig:Number_of_Waves}. For regions~F,~G and~H, the incident frequency is relatively close to the plasma frequency of the first medium, while for regions~I and~J, $\omega_{\text{i}}$ may lie relatively far from $\omega_{\text{p},1}$.
            
            \pati{Regions~F--H}{
            }
            
            Regions~F,~G and~H occur for a contradirectional moving interface ($v_{\text{m}} < 0$). Starting from region~H, scattering consists of a reflected ($\psi_{1}^{-(1)}$) and a later-backward ($\psi_{2}^{-(1)}$) wave. No later-forward wave ($\psi_{2}^{+(1)}$) is present because the transition line does not intersect with the dispersion curve of the second medium with a positive slope, so no physical forward mode in the second medium is allowed. As the magnitude of the velocity is decreased---moving from region~H to region~G---the transition line couples with a negative slope intersection point of the dispersion diagram of the second medium. This give rise to a second later-backward wave ($\psi_{2}^{-(2)}$), in addition to the reflected and first later-backward wave. Notably, the group velocity of this new wave is very small, as indicated by the nearly horizontal slope of the dispersion curve at the intersection point. This suggests that the wave remains localized near its scattering position, effectively forming a quasi-stationary mode bound to the vicinity of its scattering position. Further decreasing the velocity into region~F, the reflected wave is eliminated by the group-velocity restriction, as its trajectory is now overtaken by the interface. 
    
            \pati{Regions~H and~I}{
            }

            Regions~I and~J are particularly interesting because they can exist even when the incident frequency is not close to the plasma frequency of the first medium. These regions host a forward wave in the first medium ($\psi_{1}^{+(1)}$), as the transition line intersects the dispersion curve of the first medium with a positive slope. This necessarily implies that the group velocity of this mode is smaller than the modulation velocity; otherwise, it would be excluded by the group-velocity restriction. In region~I, the transmitted waves turn out to have complex frequencies and wavenumbers that violate the passivity constraints, and are therefore ruled out, leaving only reflected and forward waves in the first medium. 
            
            \pati{General Closing Statement}{
            }
            
            Overall, the dispersion-mediated space-time modes in Fig.~\ref{fig:Number_of_Waves} emerge because the dispersion curves allow multiple intersections with the transition line. Each additional intersection corresponds to a new scattered mode, provided it satisfies group-velocity and passivity constraints (Tab~\ref{tab:Summary_Conditions_Frequency_Transitions}). These additional modes do not appear in the nondispersive limit, where the dispersion curves are linear and intersect the transition line at most once per medium and direction.

    \section{Application to Lorentz Media}\label{sec:Application_to_Lorentz_Media}
        \pati{Lorentz Dispersive Model}{
        }

\begin{figure}[h!]
            \centering
            \includegraphics[width=0.6\linewidth]{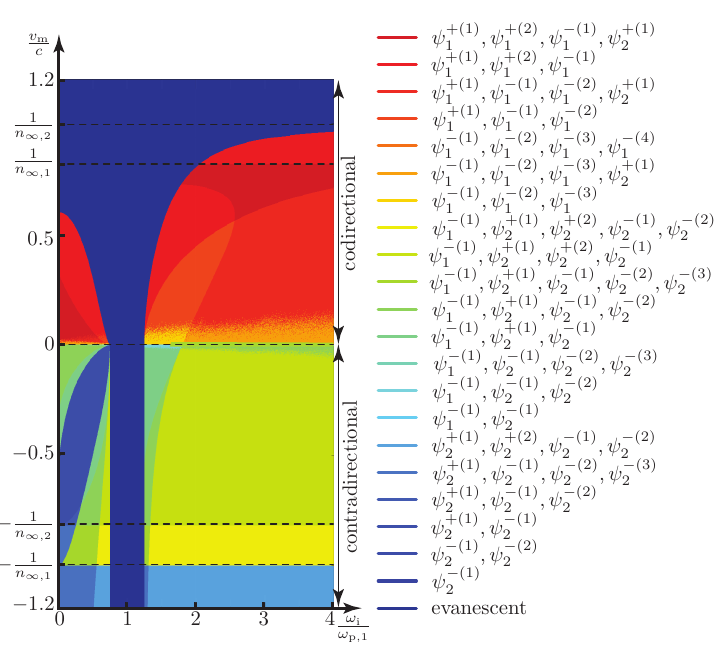}
            \caption{Dispersion-mediated space-time states for the lossless Lorentz dispersion model with parameters $n_{\infty,1} = 1, \omega_{\text{p},1}=4, \omega_{0,1} =3, n_{\infty,2} = 1.5, \omega_{\text{p},2} = 8$ and $\omega_{0,2} = 3$, computed according to the conditions in Tab.~\ref{tab:Summary_Conditions_Frequency_Transitions}. The dashed lines indicate the nondispersive velocity limits.}
            \label{fig:Number_of_Waves_Lorentz}
        \end{figure}

        The methodology presented in Sec.~\ref{sec:Frequency_Transitions} is applicable to any dispersive model, such as e.g. the Lorentz dispersive model, where $n{\left[\omega\right]} = \sqrt{n_{\infty}^{2}+\omega_{\text{p}}^{2}/\left(\omega_{0}^{2}-i\gamma\omega-\omega^{2}\right)}$, where $\omega_{\text{p}}$, $\omega_{0}$ and $\gamma$ are the plasma frequency, resonance frequency and damping factor, respectively. In this case, the characteristic equations [Eq.~\eqref{eq:General_Characterstic_Equations}] become fourth-order polynomials, potentially yielding up to four distinct solutions and thus considerably more complex and richer scattering results. Figure~\ref{fig:Number_of_Waves_Lorentz} shows the different regions for a temporal Lorentz dispersion example, allowing up to $21$ regimes. As before, some nondispersive scattering regimes reappear, but are complemented by dispersion-mediated space-time states, which gradually reduce to the nondispersive behavior at high frequencies. These Lorentzian space-time states further demonstrate that dispersion not only perturbs nondispersive scattering but can fundamentally restructure the scattering landscape, creating new classes of modes and transitions that disappear in the nondispersive limit.

    \section{Scattering Coefficients}\label{sec:Scattering_Coefficients}
        \pati{Failure of Nondispersive Methods}{
        }

        While the previous sections restricted to the frequency transitions of the scattered waves, a complete description of the electromagnetic scattering problem also requires their amplitudes. These amplitudes are determined by enforcing boundary conditions at the moving interface, which for uniform motion reduce to the continuity of $E{\left[z,t\right]} - v_{\text{m}}B{\left[z,t\right]}$ and $H{\left[z,t\right]}-v_{\text{m}}D{\left[z,t\right]}$~\cite{Caloz2019b_ST_Metamaterials_PUB}. Evaluating these conditions requires expressing the induced fields $B{\left[z,t\right]}$ and $D{\left[z,t\right]}$ in terms of the driving fields $H{\left[z,t\right]}$ and $E{\left[z,t\right]}$, which are connected through the constitutive relations of the medium. In nondispersive systems, the constitutive relations are instantaneous in time, $B{\left[z,t\right]} = n\eta/c H{\left[z,t\right]}$ and $D{\left[z,t\right]} = n/\left(c\eta\right)E{\left[z,t\right]}$, allowing the induced fields to be directly substituted into the time-domain boundary conditions, which makes the determination of the scattering amplitudes relatively straightforward. However, in dispersive media, causality requires that the response at any time $t$ depends on the entire history of the induced fields, so that, in time domain, the constitutive relations take the form of convolutions. As a result, the time-domain boundary conditions involve convolutions, coupling the present values of the fields at the interface to all past times. This coupling renders nondispersive methods challenging, because one would have to track the entire temporal evolution of the fields to evaluate the boundary conditions at each instant. Transforming to the frequency domain resolves this issue, as, in Fourier space, the convolutions in the constitutive relations become simple algebraic products, $\tilde{B}{\left[z,\omega\right]} = n{\left[\omega\right]}\eta{\left[\omega\right]}/c \tilde{H}{\left[z,\omega\right]}$ and $\tilde{D}{\left[z,\omega\right]} = n{\left[\omega\right]}/\left(c\eta{\left[\omega\right]}\right) \tilde{E}{\left[z,\omega\right]}$, so the material response at each frequency can be handled independently. This observation motivates a Fourier approach to solving dispersive scattering at moving interfaces.
        
        \pati{Approach}{
        }

        The moving interface and the dispersive medium play complementary roles in the scattering problem. On one hand, the interface motion is evaluated in the \emph{time} domain, which determines the location where the boundary conditions are applied and governs the frequency transitions of the scattered waves (Sec.~\ref{sec:Frequency_Transitions}). On the other hand, the dispersive response of the medium is treated in the \emph{frequency} domain, where the temporal convolutions in the constitutive relations reduce to algebraic products for each frequency component. This separation allows the scattering problem to be solved efficiently: the interface kinematics are enforced directly in time, while the material response is handled through its frequency-dependent properties, without requiring explicit tracking of the full temporal evolution of the fields. Together, these representations reduce the boundary-value problem to a set of algebraic equations for the spectral amplitudes of the reflected and transmitted waves.
        
        \pati{Results}{
        }

        To illustrate this approach, we consider the subluminal scattering region (region~C in Fig~\ref{fig:Number_of_Waves}), where an incident wave, $\psi_{\text{i}}$, generates one reflected wave, $\psi_{1}^{-}$, and one transmitted wave, $\psi_{2}^{+}$. For a monochromatic incident field with frequency $\omega_{\text{i}}$, the Fourier scattered waves are given by (Sec.~\ref{sec:appendix:Scattering_Coefficients})
        \begin{subequations}\label{eq:Fourier_Scattered_Waves}
            \begin{align}
                \tilde{\psi}_{1}^{-}\left[\omega\right] &= \frac{\eta_{1}{\left[\omega_{1}^{-}\right]}}{\eta_{1}{\left[\omega_{\text{i}}\right]}}\frac{\eta_{2}{\left[\omega_{2}^{+}\right]}-\eta_{1}{\left[\omega_{\text{i}}\right]}}{\eta_{2}{\left[\omega_{2}^{+}\right]}+\eta_{1}{\left[\omega_{1}^{-}\right]}}\tilde{\psi}_{\text{i}}\left[\frac{1+n_{1}{\left[\omega_{1}^{-}\right]}v_{\text{m}}/c}{1- n_{1}{\left[\omega_{\text{i}}\right]}v_{\text{m}}/c}\omega\right]\,, \\
                \tilde{\psi}_{2}^{+}\left[\omega\right] &= \frac{\eta_{2}{\left[\omega_{2}^{+}\right]}}{\eta_{1}{\left[\omega_{\text{i}}\right]}}\frac{\eta_{1}{\left[\omega_{1}^{-}\right]}+\eta_{1}{\left[\omega_{\text{i}}\right]}}{\eta_{2}{\left[\omega_{2}^{+}\right]}+\eta_{1}{\left[\omega_{1}^{-}\right]}}\tilde{\psi}_{\text{i}}\left[\frac{1-n_{2}{\left[\omega_{2}^{+}\right]}v_{\text{m}}/c}{1-n_{1}{\left[\omega_{\text{i}}\right]}v_{\text{m}}/c}\omega\right]\,,
            \end{align}
        \end{subequations}
        where $\omega_{1}^{-}$ and $\omega_{2}^{+}$ are determined by the phase matching equations [Eqs.~\eqref{eq:General_Characterstic_Equations}] and depend on the incident frequency. For a Drude model, these expressions are given in Eqs.~\eqref{eq:Frequency_Transitions_Example}. The prefactors in Eqs.~\eqref{eq:Fourier_Scattered_Waves} describe the impedance mismatch, whereas the factors multiplying $\omega$ encode the inverse Doppler frequency shift generated by the interface motion. Equations~\eqref{eq:Fourier_Scattered_Waves} have the same functional form as the reflection and transmission coefficients for a nondispersive interface~\cite{Deck-Leger2019_Uni_Vel_PUB}, but in dispersive media the coefficients are frequency-dependent, ultimately depending on the incident frequency. In the limit of nonmagnetic media, the coefficients reduce to the results of~\cite{Agrawal2021_Refl_and_Trans_USTEM_PUB}, which were obtained using the slowly-varying envelope equation.

        \pati{Other Regions}{
        }

        Equations~\eqref{eq:Fourier_Scattered_Waves} are valid only in the subluminal regime, where exactly one reflected wave and one transmitted wave is generated. In other regimes that also produce two scattered waves, the scattering solutions are generally different, but can be obtained in a similar manner using the same methodology. In scenarios involving more than two scattered waves (e.g., region~G in Fig.~\ref{fig:Number_of_Waves}), additional boundary conditions are required to fully determine the amplitudes. This situation occurs for instance in the interluminal scattering regime in nondispersive systems, where an extra condition is also necessary~\cite{Ostrovskii1967_Inter_PUB}. While several approaches have been proposed for handling additional boundary conditions in pure-time systems~\cite{Engheta2021_Dispersion_PUB,Horsley2025_Sym_Modes_Disp_TEM_PUB,Horsley2025_Scat_TEM_Drude_Lorentz_TEM}, no general additional boundary conditions has been established for space–time varying systems. A systematic treatment of these cases is left for future work.

    \section{Pulses}\label{sec:Pulses}
        
        \pati{Reflected and Transmitted Pulses}{
        }
        
        The analysis of the previous section focused on monochromatic waves and determined the scattering coefficients for a single frequency component [Eqs.~\eqref{eq:Fourier_Scattered_Waves}]. This framework can be extended to study the scattering of an incident pulse by exploiting the linearity of Maxwell’s equations and of the boundary conditions at the moving interface. Consider an incident pulse, $\psi_{\text{i}}{\left[z,t\right]}$, with a spectrum, $\tilde{\psi}_{\text{i}}{\left[\omega\right]}$, 
        \begin{equation}\label{eq:Incident_Wave_as_Fourier_Transform}
            \psi_{\text{i}}{\left[z,t\right]} = \frac{1}{2\pi}\int \dd{\omega} \tilde{\psi}_{\text{i}}{\left[\omega\right]}e^{i\left(\beta_{1}{\left[\omega\right]}z-\omega t\right)}\,.
        \end{equation}
        Each frequency component in Eq.~\eqref{eq:Incident_Wave_as_Fourier_Transform}, $\omega$, propagates independently and is scattered according to the monochromatic solutions derived in Sec.~\ref{sec:Scattering_Coefficients} [Eqs.~\eqref{eq:Fourier_Scattered_Waves}]. In other words, each frequency can be treated as a fixed monochromatic wave and the reflected and transmitted pulses can be reconstructed by inverse Fourier transform of the corresponding monochromatic scattered fields, i.e.,
        \begin{subequations}\label{eq:Pulses_Time_Domain_Solution}
            \begin{align}
                \psi_{1}^{-}{\left[z,t\right]} &= \frac{1}{2\pi}\int \dd{\omega} \tilde{\psi}_{1}^{-}{\left[\omega\right]}e^{i\left(\beta_{1}{\left[\omega\right]}z-\omega t\right)} \nonumber \\
                &= \frac{1}{2\pi}\int \dd{\omega} \frac{\eta_{1}{\left[\omega_{1}^{-}\right]}}{\eta_{1}{\left[\omega\right]}}\frac{\eta_{2}{\left[\omega_{2}^{+}\right]}-\eta_{1}{\left[\omega\right]}}{\eta_{2}{\left[\omega_{2}^{+}\right]}+\eta_{1}{\left[\omega_{1}^{-}\right]}}e^{i\left(\beta_{1}{\left[\omega\right]}z-\omega t\right)} \tilde{\psi}_{\text{i}}\left[\frac{1+n_{1}{\left[\omega_{1}^{-}\right]}v_{\text{m}}/c}{1- n_{1}{\left[\omega\right]}v_{\text{m}}/c}\omega\right]\,, \\
                \psi_{2}^{+}{\left[z,t\right]} &= \frac{1}{2\pi}\int \dd{\omega} \tilde{\psi}_{2}^{+}{\left[\omega\right]}e^{i\left(\beta_{2}{\left[\omega\right]}z-\omega t\right)} \nonumber \\
                &= \frac{1}{2\pi}\int \dd{\omega} \frac{\eta_{2}{\left[\omega_{2}^{+}\right]}}{\eta_{1}{\left[\omega\right]}}\frac{\eta_{1}{\left[\omega_{1}^{-}\right]}+\eta_{1}{\left[\omega\right]}}{\eta_{1}{\left[\omega_{1}^{-}\right]}+\eta_{2}{\left[\omega_{2}^{+}\right]}}e^{i\left(\beta_{2}{\left[\omega\right]}z-\omega t\right)} \tilde{\psi}_{\text{i}}\left[\frac{1-n_{2}{\left[\omega_{2}^{+}\right]}v_{\text{m}}/c}{1-n_{1}{\left[\omega\right]}v_{\text{m}}/c}\omega\right]\,,
            \end{align}
        \end{subequations}
        where $\tilde{\psi}_{1}^{-}{\left[\omega\right]}$ and $\tilde{\psi}_{2}^{+}{\left[\omega\right]}$ are obtained from the monochromatic scattering coefficients [Eqs.~\eqref{eq:Fourier_Scattered_Waves}] and $\omega_{1}^{-}$ and $\omega_{2}^{+}$ are functions of $\omega$ determined from the characteristic equations [Eqs.~\eqref{eq:General_Characterstic_Equations}]. In general, these integrals must be evaluated numerically to reconstruct the full pulse shapes.

        \pati{Dispersion Diagram}{
        }

        Unlike the monochromatic case, where the scattering behavior is fixed by a single frequency point, a broadband pulse samples a range of frequencies that may fall into different scattering regimes (Fig.~\ref{fig:Number_of_Waves}). As a result, distinct spectral components can undergo qualitatively different frequency transitions and their amplitudes change upon interaction with the moving interface. When recombined in the time domain, these differences may lead to substantial temporal and spectral distortion of the reflected and transmitted pulses.

        \pati{Example}{
        \begin{itemize}
            \item Perhaps make figure with the following:
            \item (a)~Dispersion diagram with multiple frequencies with center frequency in subluminal but stretches out to region H, or even D. 
            \item (b)~Plot of the scattering coefficients as a function of $\omega_{\text{i}}$.
        \end{itemize}
        }

    \section{Conclusions}\label{sec:Conclusions}
        \pati{Conclusions}{
        }
        
        We have presented a general methodology for determining the spectral content of waves scattered at a uniformly moving interface between dispersive media. By combining phase-matching requirements with two conditions---group-velocity restriction and media passivity---we have identified the physically allowed frequency transitions among the multiple solutions of the characteristic equations. A key feature of this methodology is that it is valid for arbitrary dispersion profiles, including regions near resonances, without relying on a Taylor expansion of the dispersion relation. It also does not require frame hopping, which makes it suitable for systems with multiple velocities~\cite{Deck-Leger2019_Uni_Vel_PUB,Bahrami2025_Wedges_USTEM_PUB,Luo2025_ST_Super_Focu_USTEM_PUB} or accelerated interfaces~\cite{DeKinder2025_Scat_Chirp_ASTEM}. Applying this framework to the lossless Drude model, we have showed that, in addition to the conventional nondispersive modes, a variety of dispersion-mediated space-time modes emerge. These modes have no counterpart in nondispersive scattering problems and arise from additional intersections between the dispersion curves and the phase-matching transition line. Similar behavior is observed in Lorentz media, where the higher-order characteristic equations and resonant dispersion yield an even richer set of space-time states, with up to $21$ distinct scattering regimes. Beyond the frequency transitions, we have derived the scattering coefficients for dispersive moving interfaces in the subluminal regime and formulated a Fourier-domain approach that yields the complete electromagnetic scattering solution for both monochromatic waves and broadband pulses. This formulation consistently combines interface kinematics with the causal, frequency-dependent material response and reveals how different spectral components of a pulse may undergo distinct scattering processes, leading to strong temporal and spectral reshaping. These results may provide a foundational framework for analyzing realistic space-time varying systems, especially those relying on strongly dispersive materials such as epsilon-near-zero platforms in the optical regime~\cite{Boyd2020_Time_Refr_ENZ_TEM_PUB,Segev2023_Single_Cycle_TEM_PUB,Kinsey2025_ST_Knife_TEM_PUB}. By explicitly accounting for dispersion and its role in creating new frequency-transition branches, our approach offers a systematic route to dispersion-engineer space-time metamaterials. 
    
        \pati{Future Work}{
        }

        While the present work provides a complete solution in regimes supporting two scattered waves, scattering scenarios involving a larger number of modes require additional boundary conditions. Identifying a systematic and physically grounded set of such conditions for general space-time varying dispersive systems remains an open problem and represents an important direction for future research. Other promising avenues for future research include extending the presented methodology to accelerated interfaces or multiple-velocity systems. 

        \textbf{Acknowledgments} \\
        K.D.K. is supported by the Research Foundation -- Flanders (FWO) doctoral fellowship 1174526N.

    \appendix
    \section{Scattering Coefficients}\label{sec:appendix:Scattering_Coefficients}
        \pati{Assumptions}{
        }
        
        This section derives the electromagnetic scattering solutions at a constant-velocity moving modulation interface in the subluminal regime (region~C in Fig.~\ref{fig:Number_of_Waves}). The interface separates two isotropic, linear but dispersive media, characterized by frequency-dependent refractive indices $n_{1,2}{\left[\omega\right]}$ and impedances $\eta_{1,2}{\left[\omega\right]}$. The interface follows the trajectory $z{\left[t\right]} = v_{\text{m}}t$, parametrized with laboratory time,~$t$.

        \pati{Traveling Wave Form}{
        }
        
        We consider a one-dimensional geometry in which the interface moves along the $z$-direction, the electric field is polarized along the $x$-direction and magnetic field lies along the $y$-direction. We further consider a forward-propagating monochromatic incident wave with frequency $\omega_{\text{i}}$ in the first medium, which can be written as
        \begin{equation}\label{eq:appendix:General_Forward_Incident_Wave}
            \exp\left(i\left(\beta_{1}{\left[\omega_{\text{i}}\right]}z-\omega_{\text{i}}t\right)\right) = \exp\left(i\omega_{\text{i}}\left(n_{1}{\left[\omega_{\text{i}}\right]}\frac{z}{c}-t\right)\right)\,,
        \end{equation}
        where $\beta_{1}{\left[\omega_{\text{i}}\right]} = n_{1}{\left[\omega_{\text{i}}\right]}\omega_{\text{i}}/c$. Eq.~\eqref{eq:appendix:General_Forward_Incident_Wave} defines a natural traveling-wave variable $\xi = n_{1}{\left[\omega_{\text{i}}\right]}z/c-t$, with phase velocity $v_{\text{p}} = c/n_{1}{\left[\omega_{\text{i}}\right]}$. Therefore, we propose as ansatz that all scattered waves can be expressed in the general form $\psi_{i}^{\pm}{\left[\xi_{i}^{\pm}\right]} = \exp\left(i\omega_{i}^{\pm}\xi_{i}^{\pm}\right)$, where $\xi_{i}^{\pm} = \pm\left(n_{i}{\left[\omega_{i}^{\pm}\right]}z/c\mp t\right)$, $i$ denotes the medium and the superscript $\pm$ denote forward ($+$) propagating waves and backward ($-$) propagating waves. 

        \pati{General Traveling Waves Subluminal}{
        }

        In the subluminal regime, an incident wave generates a reflected wave ($\psi_{1}^{-}$), propagating backward in the first medium and a transmitted wave ($\psi_{2}^{+}$), propagating forward in the second medium. The electric fields may be written as
        \begin{subequations}\label{eq:appendix:Electric_Waveforms}
            \begin{align}
                E_{\text{i}} &= \psi_{\text{i}}{\left[n_{1}{\left[\omega_{\text{i}}\right]}\frac{z}{c}-t\right]} \,, \\
                E_{1}^{-}    &= \psi_{1}^{-}{\left[-\left(n_{1}{\left[\omega_{1}^{-}\right]}\frac{z}{c}+t\right)\right]}\,, \\
                E_{2}^{+}    &= \psi_{2}^{+}{\left[n_{2}{\left[\omega_{2}^{+}\right]}\frac{z}{c}-t\right]}\,, 
            \end{align}
        \end{subequations}
        and the magnetic fields can be written as
        \begin{subequations}\label{eq:appendix:Magnetic_Waveforms}
            \begin{align}
                H_{\text{i}} &=  \frac{1}{\eta_{1}{\left[\omega_{\text{i}}\right]}}\psi_{\text{i}}{\left[n_{1}{\left[\omega_{\text{i}}\right]}\frac{z}{c}-t\right]} \,, \\
                H_{1}^{-}    &= -\frac{1}{\eta_{1}{\left[\omega_{1}^{-}\right]}}\psi_{1}^{-}{\left[-\left(n_{1}{\left[\omega_{1}^{-}\right]}\frac{z}{c}+t\right)\right]}\,, \\
                H_{2}^{+}    &=  \frac{1}{\eta_{2}{\left[\omega_{2}^{+}\right]}}\psi_{2}^{+}{\left[n_{2}{\left[\omega_{2}^{+}\right]}\frac{z}{c}-t\right]}\,, 
            \end{align}
        \end{subequations}
        where $\psi_{\text{i}}$, $\psi_{1}^{-}$ and $\psi_{2}^{+}$ denote the incident, reflected and transmitted waves, respectively, and $\omega_{\text{i}}$, $\omega_{1}^{-}$ and $\omega_{2}^{+}$ are the corresponding frequencies~\footnote{We have dropped the $(j)$ superscript used in Sec.~\ref{sec:Frequency_Transitions} to distinguish different waves with same features as there is only one reflected and one transmitted wave in the subluminal regime (region~C in Fig.~\ref{fig:Number_of_Waves}).}. The reflected and transmitted frequencies are determined by the general phase matching conditions [Eqs.~\eqref{eq:General_Characterstic_Equations}] or by their particular counterparts of Eqs.~\eqref{eq:Frequency_Transitions_Example} for the Drude-model example [Eq.~\eqref{eq:Refractive_Index_Profile_Free_Electron_Plasma_Model}].

        \pati{Problem of Time Boundary Conditions}{
        }
        
        The boundary conditions are the continuity of $E{\left[z,t\right]} - v_{\text{m}}B{\left[z,t\right]}$ and $H{\left[z,t\right]}-v_{\text{m}}D{\left[z,t\right]}$ at the moving interface. They require knowing $D{\left[z,t\right]}$ and $B{\left[z,t\right]}$, which are related to $E{\left[z,t\right]}$ and $H{\left[z,t\right]}$ via the constitutive relations. In nondispersive media, these relations are simple products: $B{\left[z,t\right]} = n\eta/c H{\left[z,t\right]}$ and $D{\left[z,t\right]} = n/\left(c\eta\right)E{\left[z,t\right]}$. In dispersive media, the constitutive relations become convolutions in time, which are cumbersome to handle~\footnote{In the monochromatic case of Eq.~\eqref{eq:appendix:General_Forward_Incident_Wave}, these relations become once again simple products, yet we formulate the analysis in the Fourier domain because this approach extends directly to pulses (Sec.~\ref{sec:Pulses}), which are composed of multiple frequency components.}. However, in Fourier domain, these constitutive relations reduce again to simple frequency-domain products: $\tilde{B}{\left[z,\omega\right]} = n{\left[\omega\right]}\eta{\left[\omega\right]}/c \tilde{H}{\left[z,\omega\right]}$ and $\tilde{D}{\left[z,\omega\right]} = n{\left[\omega\right]}/\left(c\eta{\left[\omega\right]}\right)\tilde{E}{\left[z,\omega\right]}$.

        \pati{Boundary Conditions}{
        }
        
        The time-domain boundary conditions at $z=v_{\text{m}}t$ are given by
        \begin{subequations}
            \begin{align}
                &\left.\left(E_{\text{i}}{\left[n_{1}{\left[\omega_{\text{i}}\right]}\frac{z}{c} - t\right]} - v_{\text{m}}B_{\text{i}}{\left[n_{1}{\left[\omega_{\text{i}}\right]}\frac{z}{c} - t\right]}\right) + \left(E_{1}^{-}{\left[-\left(n_{1}{\left[\omega_{1}^{-}\right]}\frac{z}{c} + t\right)\right]} - v_{\text{m}}B_{1}^{-}{\left[-\left(n_{1}{\left[\omega_{1}^{-}\right]}\frac{z}{c} + t\right)\right]}\right)\right|_{z=v_{\text{m}}t} \nonumber\\
                &\hspace{8cm}= \left.\left(E_{2}^{+}{\left[n_{2}{\left[\omega_{2}^{+}\right]}\frac{z}{c} - t\right]} - v_{\text{m}}B_{2}^{+}{\left[n_{2}{\left[\omega_{2}^{+}\right]}\frac{z}{c} - t\right]}\right)\right|_{z=v_{\text{m}}t} \,, \\
                &\left.\left(H_{\text{i}}{\left[n_{1}{\left[\omega_{\text{i}}\right]}\frac{z}{c} - t\right]} - v_{\text{m}}D_{\text{i}}{\left[n_{1}{\left[\omega_{\text{i}}\right]}\frac{z}{c} - t\right]}\right) + \left(H_{1}^{-}{\left[-\left(n_{1}{\left[\omega_{1}^{-}\right]}\frac{z}{c} + t\right)\right]} - v_{\text{m}}D_{1}^{-}{\left[-\left(n_{1}{\left[\omega_{1}^{-}\right]}\frac{z}{c} + t\right)\right]}\right)\right|_{z=v_{\text{m}}t} \nonumber \\
                &\hspace{8cm}= \left.\left(H_{2}^{+}{\left[n_{2}{\left[\omega_{2}^{+}\right]}\frac{z}{c} - t\right]} - v_{\text{m}}D_{2}^{+}{\left[n_{2}{\left[\omega_{2}^{+}\right]}\frac{z}{c} - t\right]}\right)\right|_{z=v_{\text{m}}t} \,,
            \end{align}
        \end{subequations}
        where all fields are evaluated at the interface position. Incorporating the interface trajectory yields
        \begin{subequations}\label{eq:appendix:System_of_Equations_Subluminal}
            \begin{align}
                &E_{\text{i}}{\left[-\left(1 - n_{1}{\left[\omega_{\text{i}}\right]}\frac{v_{\text{m}}}{c}\right)t\right]} - v_{\text{m}}B_{\text{i}}{\left[-\left(1 - n_{1}{\left[\omega_{\text{i}}\right]}\frac{v_{\text{m}}}{c}\right)t\right]} + E_{1}^{-}{\left[-\left(1+n_{1}{\left[\omega_{1}^{-}\right]}\frac{v_{\text{m}}}{c}\right)t\right]} - v_{\text{m}}B_{1}^{-}{\left[-\left(1 + n_{1}{\left[\omega_{1}^{-}\right]}\frac{v_{\text{m}}}{c}\right)t\right]} \nonumber \\
                &\hspace{7cm}= E_{2}^{+}{\left[-\left(1 - n_{2}{\left[\omega_{2}^{+}\right]}\frac{v_{\text{m}}}{c}\right)t\right]} - v_{\text{m}}B_{2}^{+}{\left[-\left(1 - n_{2}{\left[\omega_{2}^{+}\right]}\frac{v_{\text{m}}}{c}\right)t\right]}\,, \\
                &H_{\text{i}}{\left[-\left(1 - n_{1}{\left[\omega_{\text{i}}\right]}\frac{v_{\text{m}}}{c}\right)t\right]} - v_{\text{m}}D_{\text{i}}{\left[-\left(1 - n_{1}{\left[\omega_{\text{i}}\right]}\frac{v_{\text{m}}}{c}\right)t\right]} + H_{1}^{-}{\left[-\left(1 + n_{1}{\left[\omega_{1}^{-}\right]}\frac{v_{\text{m}}}{c}\right)t\right]} - v_{\text{m}}D_{1}^{-}{\left[-\left(1 + n_{1}{\left[\omega_{1}^{-}\right]}\frac{v_{\text{m}}}{c}\right)t\right]} \nonumber \\
                &\hspace{7cm}= H_{2}^{+}{\left[-\left(1 - n_{2}{\left[\omega_{2}^{+}\right]}\frac{v_{\text{m}}}{c}\right)t\right]} - v_{\text{m}}D_{2}^{+}{\left[-\left(1 - n_{2}{\left[\omega_{2}^{+}\right]}\frac{v_{\text{m}}}{c}\right)t\right]}\,.
            \end{align}
        \end{subequations}
        Now comes the central idea of the proposed method. We Fourier-transform Eqs.~\eqref{eq:appendix:System_of_Equations_Subluminal}, while including the frequency-domain constitutive relations, inserting the general waveforms Eqs.~\eqref{eq:appendix:Electric_Waveforms} and Eqs.~\eqref{eq:appendix:Magnetic_Waveforms}, and using the Fourier scaling property
        \begin{equation}
            \mathcal{F}\left\{f{\left[at\right]}\right\}{\left[\omega\right]} = \frac{1}{\left|a\right|}\tilde{f}{\left[\frac{\omega}{a}\right]}\,.
        \end{equation}
        This leads to the following simple system of equations for the reflected and transmitted waves:
        \begin{subequations}\label{eq:appendix:Fourier_System_of_Equations_Subluminal}
            \begin{align}
                \tilde{\psi}_{\text{i}}\left[-\frac{\omega}{1- n_{1}{\left[\omega_{\text{i}}\right]}v_{\text{m}}/c}\right] + \tilde{\psi}_{1}^{-}\left[-\frac{\omega}{1+ n_{1}{\left[\omega_{1}^{-}\right]}v_{\text{m}}/c}\right] &= \tilde{\psi}_{2}^{+}\left[-\frac{\omega}{1 - n_{2}{\left[\omega_{2}^{+}\right]}v_{\text{m}}/c}\right] \,, \\
                \frac{1}{\eta_{1}{\left[\omega_{\text{i}}\right]}}\tilde{\psi}_{\text{i}}\left[-\frac{\omega}{1- n_{1}{\left[\omega_{\text{i}}\right]}v_{\text{m}}/c}\right] -\frac{1}{\eta_{1}{\left[\omega_{1}^{-}\right]}}\tilde{\psi}_{1}^{-}\left[-\frac{\omega}{1 + n_{1}{\left[\omega_{1}^{-}\right]}v_{\text{m}}/c}\right] &= \frac{1}{\eta_{2}{\left[\omega_{2}^{+}\right]}}\tilde{\psi}_{2}^{+}\left[-\frac{\omega}{1 - n_{2}{\left[\omega_{2}^{+}\right]}v_{\text{m}}/c}\right]\,.
            \end{align}
        \end{subequations}
        Solving Eqs.~\eqref{eq:appendix:Fourier_System_of_Equations_Subluminal} for the reflected and transmitted waves yields
        \begin{subequations}\label{eq:appendix:Solution_Fourier_System_of_Equations_Subluminal}
            \begin{align}
                \tilde{\psi}_{1}^{-}\left[-\frac{\omega}{1 + n_{1}{\left[\omega_{1}^{-}\right]}v_{\text{m}}/c}\right] &= \frac{\eta_{1}{\left[\omega_{1}^{-}\right]}}{\eta_{1}{\left[\omega_{\text{i}}\right]}}\frac{\eta_{2}{\left[\omega_{2}^{+}\right]}-\eta_{1}{\left[\omega_{\text{i}}\right]}}{\eta_{2}{\left[\omega_{2}^{+}\right]}+\eta_{1}{\left[\omega_{1}^{-}\right]}}\tilde{\psi}_{\text{i}}\left[-\frac{\omega}{1 - n_{1}{\left[\omega_{\text{i}}\right]}v_{\text{m}}/c}\right]\,, \\
                \tilde{\psi}_{2}^{+}\left[-\frac{\omega}{1 - n_{2}{\left[\omega_{2}^{+}\right]}v_{\text{m}}/c}\right] &= \frac{\eta_{2}{\left[\omega_{2}^{+}\right]}}{\eta_{1}{\left[\omega_{\text{i}}\right]}}\frac{\eta_{1}{\left[\omega_{1}^{-}\right]}+\eta_{1}{\left[\omega_{\text{i}}\right]}}{\eta_{1}{\left[\omega_{1}^{-}\right]}+\eta_{2}{\left[\omega_{2}^{+}\right]}}\tilde{\psi}_{\text{i}}\left[-\frac{\omega}{1 - n_{1}{\left[\omega_{\text{i}}\right]}v_{\text{m}}/c}\right]\,.
            \end{align}
        \end{subequations}
        To express the scattered fields again as functions of the usual frequency variable, $\omega$, we finally introduce spectral variables that absorb the Doppler prefactors appearing in Eqs.~\eqref{eq:appendix:Solution_Fourier_System_of_Equations_Subluminal},
        \begin{subequations}\label{eq:appendix:Change_of_Variables_Subluminal}
            \begin{align}
                \omega &\mapsto -\left(1 + n_{1}{\left[\omega_{1}^{-}\right]}\frac{v_{\text{m}}}{c}\right)\omega\,, \\
                \omega &\mapsto -\left(1 - n_{2}{\left[\omega_{2}^{+}\right]}\frac{v_{\text{m}}}{c}\right)\omega\,.
            \end{align}
        \end{subequations}
        Inserting Eqs.~\eqref{eq:appendix:Change_of_Variables_Subluminal} into Eqs.~\eqref{eq:appendix:Solution_Fourier_System_of_Equations_Subluminal} yields the final solutions
        \begin{subequations}
            \begin{align}
                \tilde{\psi}_{1}^{-}\left[\omega\right]& = \frac{\eta_{1}{\left[\omega_{1}^{-}\right]}}{\eta_{1}{\left[\omega_{\text{i}}\right]}}\frac{\eta_{2}{\left[\omega_{2}^{+}\right]}-\eta_{1}{\left[\omega_{\text{i}}\right]}}{\eta_{2}{\left[\omega_{2}^{+}\right]}+\eta_{1}{\left[\omega_{1}^{-}\right]}}\tilde{\psi}_{\text{i}}\left[\frac{1+n_{1}{\left[\omega_{1}^{-}\right]}v_{\text{m}}/c}{1- n_{1}{\left[\omega_{\text{i}}\right]}v_{\text{m}}/c}\omega\right]\,, \\
                \tilde{\psi}_{2}^{+}\left[\omega\right] &= \frac{\eta_{2}{\left[\omega_{2}^{+}\right]}}{\eta_{1}{\left[\omega_{\text{i}}\right]}}\frac{\eta_{1}{\left[\omega_{1}^{-}\right]}+\eta_{1}{\left[\omega_{\text{i}}\right]}}{\eta_{2}{\left[\omega_{2}^{+}\right]}+\eta_{1}{\left[\omega_{1}^{-}\right]}}\tilde{\psi}_{\text{i}}\left[\frac{1-n_{2}{\left[\omega_{2}^{+}\right]}v_{\text{m}}/c}{1-n_{1}{\left[\omega_{\text{i}}\right]}v_{\text{m}}/c}\omega\right]\,.
            \end{align}
        \end{subequations}
        
\bibliography{main}
\end{document}